\documentclass[twocolumn]{revtex4}[12pt]
\usepackage{graphicx}
\begin{document}
\title{Heavy ion-acoustic rogue waves in electron-positron multi-ion plasmas}
\author{$^*$N. A. Chowdhury, A. Mannan, M. M. Hasan, and A. A. Mamun}
\address{Department of Physics, Jahangirnagar University, Savar,
Dhaka-1342, Bangladesh.\\
$^*$Email: nurealam1743phy@gmail.com}
\begin{abstract}
The nonlinear propagation of heavy-ion-acoustic (HIA) waves (HIAWs) in a four component multi-ion plasma
(containing inertial heavy negative ions and light positive ions, as well as inertialess nonextensive
electrons and positrons) has been theoretically investigated. The nonlinear  Schr\"{o}dinger (NLS)
equation is derived by employing the reductive perturbation method.
 It is found that the NLS equation leads to the modulational instability (MI) of HIAWs, and to the formation of HIA
rogue waves (HIARWs), which are due to the effects of nonlinearity and dispersion in the propagation of
HIAWs. The conditions for MI of HIAWs, and the basic properties of the generated HIARWs are identified.
It is observed that the striking features (viz. instability criteria, growth rate of MI,  amplitude
and width of HIARWs, etc.) of the HIAWs  are significantly modified by effects of
nonextensivity of electrons and positrons, ratio of light positive ion mass to heavy negative ion mass,
ratio of electron number density to light positive ion number density, and ratio of electron temperature to positron temperature, etc. The
relevancy of our present investigation to the observations in the space (viz. cometary comae and
earth's ionosphere) and laboratory (laser plasma interaction experimental devices) plasmas is pointed out.
\end{abstract}
\maketitle
\section{Lead Paragraph}
\textbf{A new electron-positron, multi-ion plasma model has been considered
to identify new features (instability criteria, growth rate of MI,  amplitude
and width, etc.) of heavy ion-acoustic rogue waves. This rogue waves are
associated with nonlinear propagation of HIAWs in which
the inertia (restoring force) is mainly provided by
the heavy negative ions (nonextensive electron and positron temperatures)
and are appeared as the solutions of NLS equation (derived here by the
reductive perturbation method) in unstable parametric regime.
The new striking features of these HIARWs are identified and are
 found to be applicable in the space and laboratories plasmas. }
\section{Introduction}
Over the last few decades, wave dynamics in electron-positron-ion (e-p-i) plasmas is one of the major
research area for the plasma physicists because of painstaking experimental observational
evidence in both space (viz. early universe \cite{Misner1973}, active galactic
nuclei \cite{Miller1987}, pulsar magnetosphere \cite{Michel1982}, and neutron
stars \cite{Michel1991}) and laboratories plasmas (laser-plasma interaction
research  \cite{Shukla1984}, semiconductor plasmas \cite{Shukla1986}, and
other magnetic confinement systems \cite{Surko1990}) revealed the existence of
nonlinear structures (viz. solitons, envelope solitons, shocks, vortices, rogue waves etc.)
in such kind of multicomponent plasmas. The existence of negative ions in the cometary
comae \cite{Chaizy1991} and earth's ionosphere \cite{Massey1976} is already established.
Plasma can contain negative ions along with positive ions. The
negative ions can appear in electronegative plasmas as a result of
elementary processes such as dissociative or nondissociative
electron attachment to neutrals \cite{Vladimirov2003,Mamun2003,Djebli2003}.
Positive and negative ion may coexist simultaneously in neutral beam sources
\cite{Bacal1979}, plasma processing reactor \cite{Gottscho1986},
and in low-temperature laboratory experiments \cite{Ichiki2002}. Similarly
positrons can be generated in modern laser plasma experiments when ultra-intense laser pulse interacts
with matter \cite{Kourakis2006,Esfandyari-Khalejahi2006}.

In case of space and laboratory plasmas, all time particles do not follow
Maxwellian distribution (which is a velocity distribution
describing the plasma particles in a thermal equilibrium
\cite{Maksivonic1997,Leubner2004,Christon2012}). Although, a
large number of authors considered that plasma components are in
thermal equilibrium but due to the some external disturbances
(e.g. wave-particle interactions, external force fields present in natural space plasma
environments, and turbulence, etc.) their
assumption is no longer valid. In space and astrophysical
environments,  the Maxwellian distribution is no longer exist when the plasma
particles move very fast compared to their thermal velocities.
Non-extensive generalization of the Boltzmann-Gibbs-Shannon
entropy, first recognized by Renyi \cite{Renyi1955} and subsequently proposed
by Tsallis \cite{Tsallis1988} has been obtained a great deal
of interest during last few decades. The nonextensive distribution is
generally denoted by q and in the q-nonextensive framework, the
one-dimensional equilibrium distribution function $f_s(v_s)$ is
given  \cite{Silva1998} by
\begin{eqnarray}
&&\hspace*{-2.4cm}f_s(v_s) = A_q \left[1-(q-1) \frac{m _s{v_s}^2}{2 k_B T_s } \right]^ \frac{1}{q-1}.\nonumber\
\end{eqnarray}
Here the normalization constant is
 \begin{eqnarray}
&&\hspace*{-.3cm}A_q =\frac{n_{s0}\Gamma(\frac{1}{1-q})}{\Gamma(\frac{1}{1-q}-\frac{1}{2})}\sqrt{\frac{m_s(1-q)}{2\pi k_B T_s}},~(for -1 < q < 1),\nonumber\
\end{eqnarray}
where $n_{s0}$, $m_s$, $v_s$,
and $T_s$ are the equilibrium number density, mass, thermal speed, and temperature of the
energetic particle species $s$. The thermal speed $v_s$ of the nonextensive particle species $s$ is
defined as  $(2k_BT_s/m_s)$, where $k_B$ is the familiar Boltzmann constant. Nonextensive
plasmas  are found in cosmological and astrophysical scenarios (viz. stellar polytropes
\cite{Plastino1993}, hadronic matter and quark-gluon plasma \cite{Gervino2012},
dark-matter halos \cite{Feron2008}, Earth’s bow-shock \cite{Asbridge1968}, magnetospheres of Jupiter and Saturn
\cite{Krimigis1983}, etc.) as well as laboratory applications like nanomaterials, micro-devices, and micro-structures
\cite{Vladimirov2004}, etc.

The propagation of wave packets in a nonlinear, dispersive medium is appeared to the modulation
of their wave amplitudes, because of the interaction between high and low frequency modes,
parametric wave coupling, the nonlinear self-interaction of the carrier wave modes, and such kind of
system is governed by the NLS equation which admits interesting rational solution named  rogue waves
(also familiar as freak waves, extreme waves, killer waves, and monster waves) solution in the modulationally unstable region.
The rogue waves were first observed in ocean \cite{Kharif2009}, now it can be seen in plasmas, atmospheric physics,
optics, stock market crashes \cite{Yan2010}, and super-fluid helium \cite{Ganshin2008}.

Recently Eslami \textit{et al.} \cite{Eslami2011} investigated the stability of IAWs in present of
q-nonextensive distributed electron and positron plasmas where they found nonextensive parameter plays a
considerable effects to modify stability conditions of the IAWs. Bacha \textit{et al.} \cite{Bacha2012} analyzed
the rogue waves may be notably affected by electron nonextensivity depending on whether the parameter $q$
is positive or negative. In multi-ion plasma rogue wave is examined by  Ei-Labany  \textit{et al.} \cite{El-Labany2011}
they found that within certain negative ion mass the rogue wave cannot propagate and  energy concentration
in rogue wave pulses largely depends on  negative ion mass and density. Jannat  \textit{et al.} \cite{Jannat2016}
studied Gardner solitons in a multi-ion plasma system with  nonextensive electrons and positrons.
The aim of the present work is, by employing reductive perturbation method a NLS equation is derived to study rogue waves in multi-ion plasma with
nonextensive electron and positron.

The paper is organized in the following fashion: The model equations are presented in Sec. III. By using reductive
perturbation technique, we derived a NLS equation which governs the slow amplitude evolution in space and time is
given in Sec. IV.  The stability of HIAWs and rogue waves are presented in Sec. V. The conclusion is provided in Sec. VI.
\section{ The Model Equations}
We consider an unmagnetized four component plasma system comprising of inertial light positive ions, heavy  negative ions,
as well as inertialess nonextensive electrons and positrons. At equilibrium, the charge neutrality condition
can be expressed as $Z_1n_{10} + n_{p0}= Z_2 n_{20}+ n_{e0}$, where $n_{s0}$ is the unperturbed number densities of the species $s$ ($s=1,2,e,p,$
for light positive ions, heavy negative ions, electrons, and positrons, respectively) and $Z_1~(Z_2)$ is the
charge state of positive light  ion (heavy negative ion). The normalized basic  equations governing the dynamics of
the  IAWs in our considered plasma system are given by
\begin{eqnarray}
&&\hspace*{-.8cm}\frac{\partial n_1}{\partial t}+\frac{\partial}{\partial x}(n_1 u_1)=0,\label{a1}\\
&&\hspace*{-.8cm}\frac{\partial u_1}{\partial t} + u_1\frac{\partial u_1}{\partial x}=-\frac{\partial \phi}{\partial x},\label{a1}\\
&&\hspace*{-.8cm}\frac{\partial n_2}{\partial t}+\frac{\partial}{\partial x}(n_2 u_2)=0, \label{a1}\\
&&\hspace*{-.8cm}\frac{\partial u_2}{\partial t} + u_2\frac{\partial u_2 }{\partial x}=\alpha \frac{\partial \phi}{\partial x},\label{a1}\\
&&\hspace*{-.8cm}\frac{\partial^2 \phi}{\partial x^2}=\mu_e n_e-\mu_p n_p- n_1+(1-\mu_e+\mu_p)n_2. \label{a3}
\end{eqnarray}
\noindent The number densities of electrons and positrons following $q-$ distribution are
\begin{eqnarray}
&&\hspace*{-3.2cm}n_e= [1+(q-1)\phi]^{\frac{(1+q)}{2(q-1)}},\nonumber\\
&&\hspace*{-3.2cm}n_p= [1-(q-1)\sigma \phi]^{\frac{(1+q)}{2(q-1)}},
\end{eqnarray}
where $q$ is the nonextensive parameter describing the degree of nonextensivity,
 i.e. $q=1$ corresponds to Maxwellian distribution, whereas $q<1$ refers to
the superextensivity, and the opposite condition $q>1$ refers to the subextensivity.
Substituting Eq. $(6)$ into Eq. $(5)$ and expanding up to third order, we get
\begin{eqnarray}
&&\hspace*{-1.5cm}\frac{\partial^2 \phi}{\partial x^2}=\mu_e-\mu_p -n_1+\lambda n_2+\gamma_1 \phi\nonumber\\
&&\hspace*{1.1cm}+\gamma_2 \phi^2+\gamma_3 \phi^3+\cdot\cdot\cdot\cdot\cdot, \label{Ch4}
\end{eqnarray}
where
\begin{eqnarray}
&&\hspace*{-1.2cm}\lambda=1-\mu_e+\mu_p, \nonumber\\
&&\hspace*{-1.2cm}\gamma_1=\frac{(\mu_e+\mu_p \sigma)(q+1)}{2}, \nonumber\\
&&\hspace*{-1.2cm}\gamma_2=\frac{(\mu_e-\mu_p\sigma^2 )(q+1)(3-q)}{8}, \nonumber\\
&&\hspace*{-1.2cm}\gamma_3=\frac{(\mu_e+ \mu_p\sigma^3)(q+1)(q-3)(3q-5)}{48},\nonumber\
\end{eqnarray}
and

~~~~~$\sigma=\frac{T_e}{T_p}$,~~~$\alpha=\frac{Z_2 m_1}{Z_1 m_2}$,~~~$\mu_e=\frac{n_{e0}}{n_{10}}$,~~~$\mu_p=\frac{n_{p0}}{n_{10}}$.

\noindent In the above equations, $n_1$ ($n_2$) is the number density of light
 positive ions (heavy negative ions) normalized by its equilibrium value
$n_{10}$ ($n_{20}$); $u_1(u_2)$ is the positive (negative) ion fluid speed
 normalized by $C_1=(Z_1 k_{B} T_{e}/m_1)^{1/2}$, and
$\phi$ is the electrostatic wave potential normalized by $k_B T_e/e$ (with
$e$ being the magnitude of an electron charge and $k_B$ is the Boltzmann constant).
$m_1$  $(m_2)$ is the rest mass of  light positive ion (heavy negative ion), respectively;
$T_e$ and $T_p$ is the temperature of  electrons and  positrons
respectively. The time and space variables are normalized by
${\omega^{-1}_{pi}}=(m_1/4\pi Z^2_1e^2 n_{10})^{1/2}$ and $\lambda_{D1}=(k_{B} T_{e}/4 \pi Z_1 e^2 n_{10})^{1/2}$, respectively.
\begin{figure*}[htp]
  \centering
  \begin{tabular}{ccc}
  \includegraphics[width=80mm]{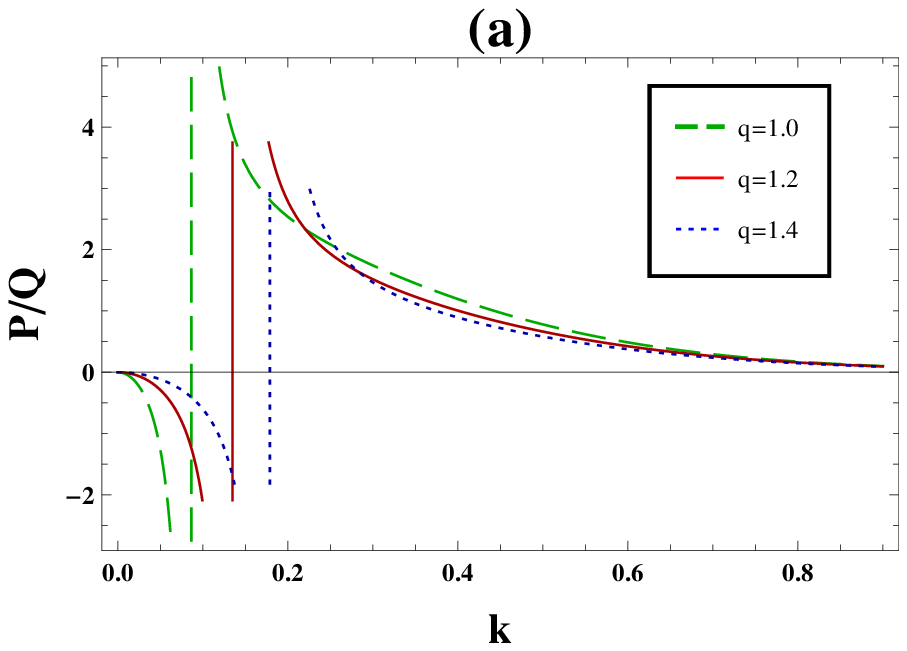}&
  \hspace{0.15in}
  \includegraphics[width=80mm]{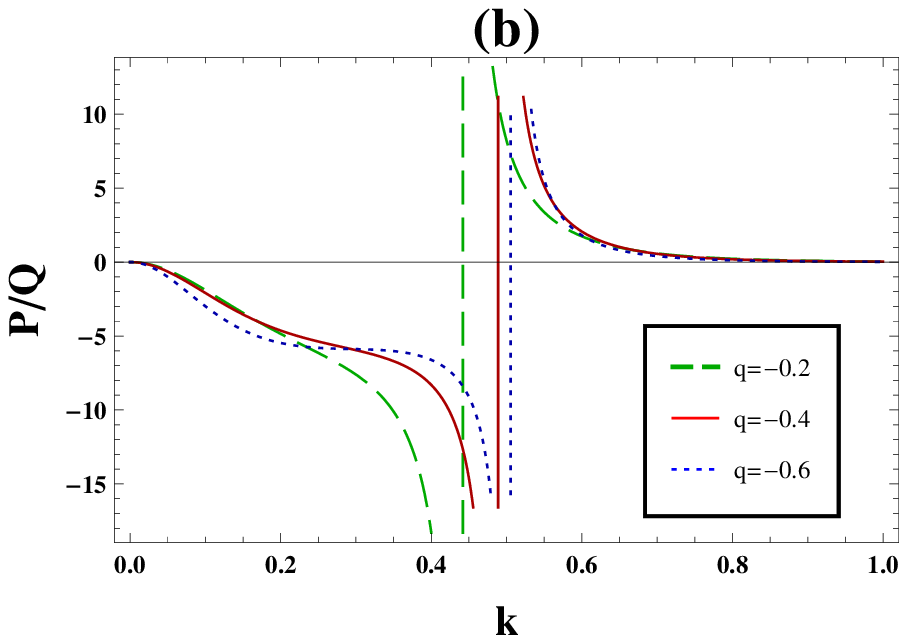}\\
  \end{tabular}
  \label{figur}\caption{Showing the variation of $P/Q$ against $k$ for different values of $q$.
  (a) For $q=$ positive. (b)  For $q=$ negative. Generally all the figures are generated by using
  these values $\alpha=0.5,\sigma=0.3,\mu_e=0.7,$ and  $\mu_p=0.5$.}
\end{figure*}
\begin{figure*}[htp]
  \centering
  \begin{tabular}{ccc}
  \includegraphics[width=80mm]{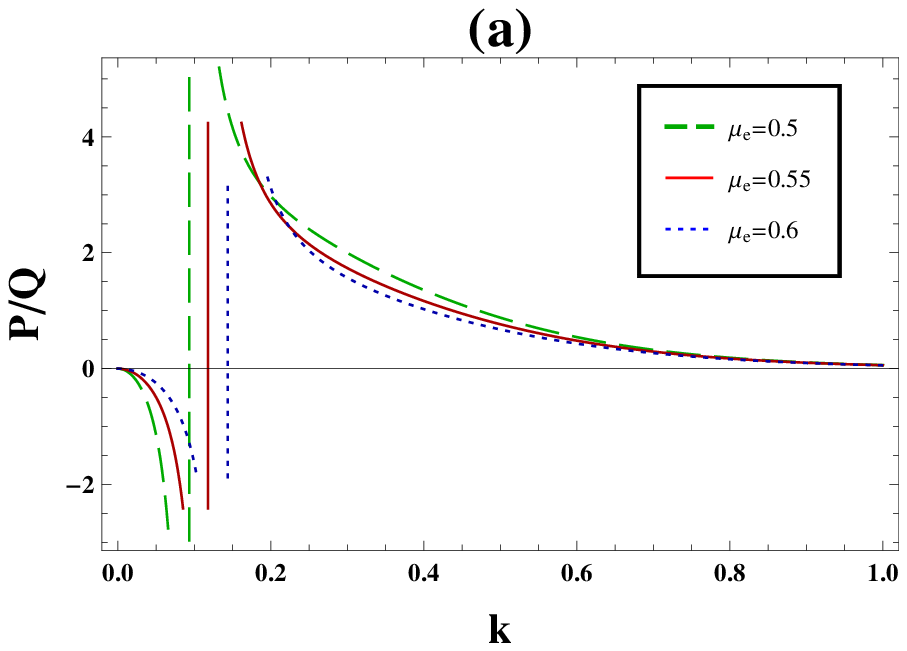}&
  \hspace{0.15in}
  \includegraphics[width=80mm]{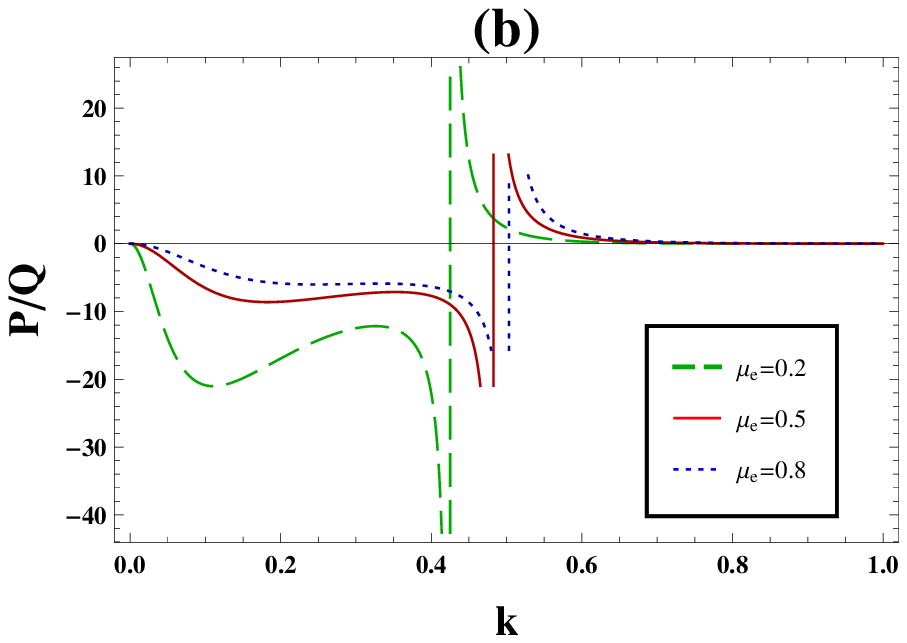}\\
  \end{tabular}
  \label{figur}\caption{Variation of $P/Q$ against $k$ for different values of $\mu_e$.
  (a) For  $q=1.5$. (b) For $q=-0.7$.}
\end{figure*}
\begin{figure*}[htp]
  \centering
  \begin{tabular}{ccc}
  \includegraphics[width=80mm]{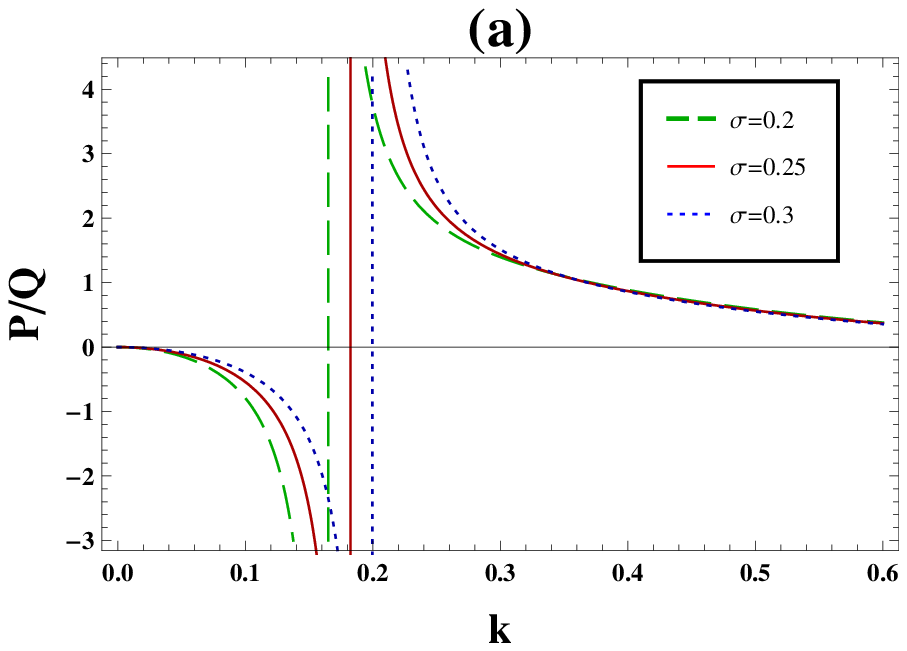}&
  \hspace{0.15in}
  \includegraphics[width=80mm]{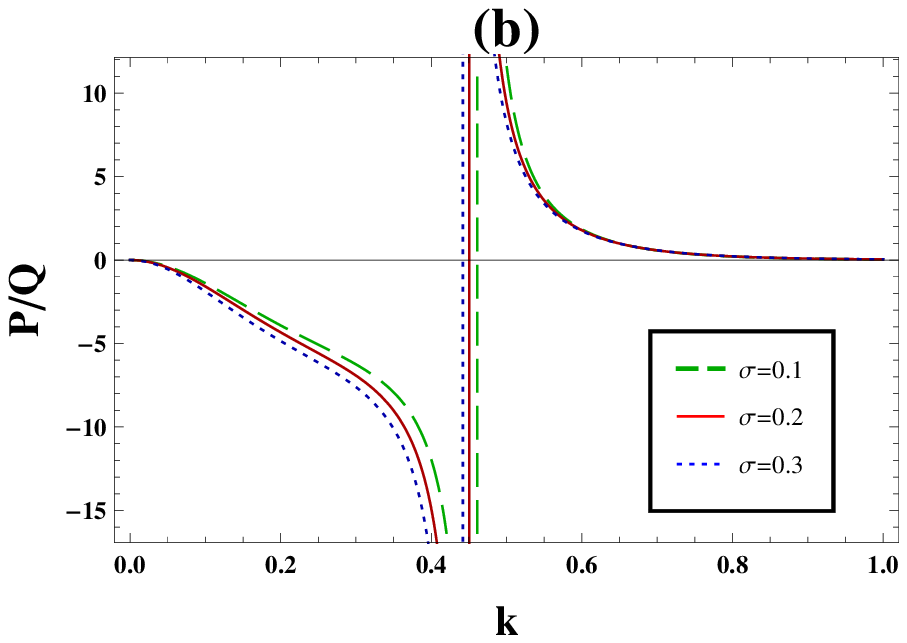}\\
  \end{tabular}
  \label{figur}\caption{Showing the variation of $P/Q$ against $k$ for different values of $\sigma$.
  (a) For $q=1.5$. (b) For  $q=-0.2$.}
\end{figure*}
\begin{figure*}[htp]
  \centering
  \begin{tabular}{ccc}
  \includegraphics[width=80mm]{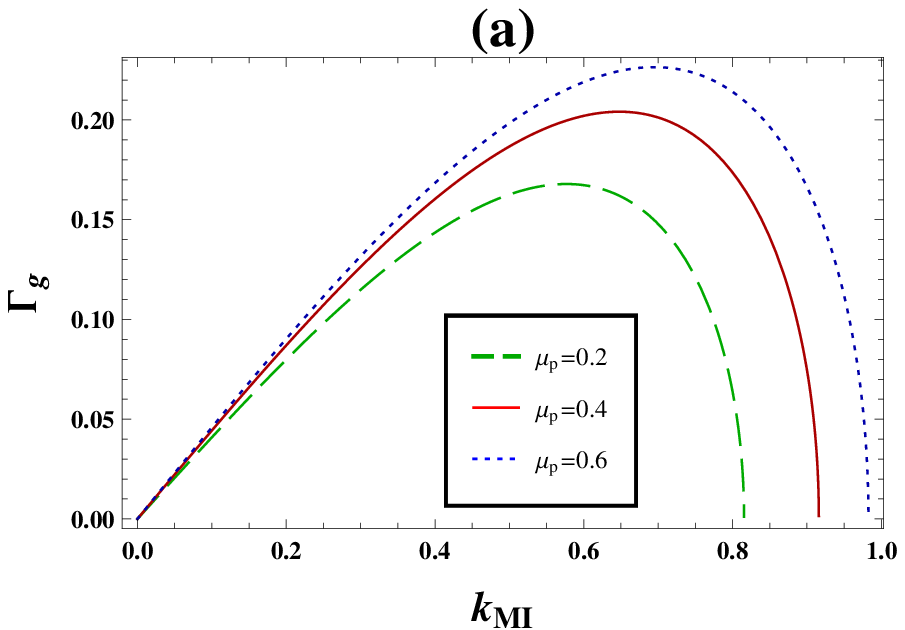}&
  \hspace{0.15in}
  \includegraphics[width=80mm]{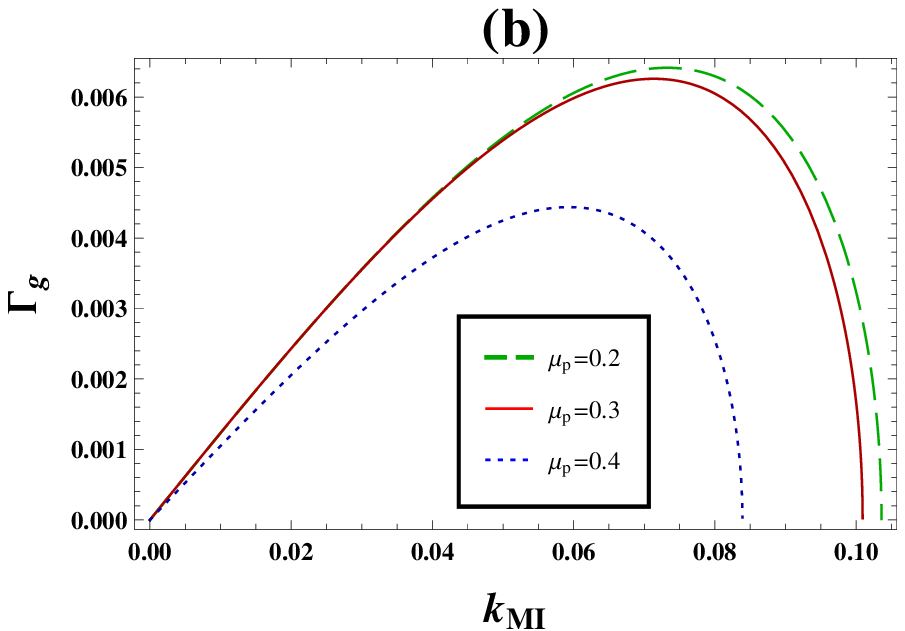}\\
  \end{tabular}
  \label{figur}\caption{Plot of the of MI growth rate $(\Gamma_g)$ against ${k_{MI}}$ for different values of $\mu_p$.
  (a) For $q=1.5$. (b) For $q=-0.7$. Along with $k=0.5$ and $\Phi_0=0.5$.}
\end{figure*}
\begin{figure*}[htp]
  \centering
  \begin{tabular}{ccc}
  \includegraphics[width=80mm]{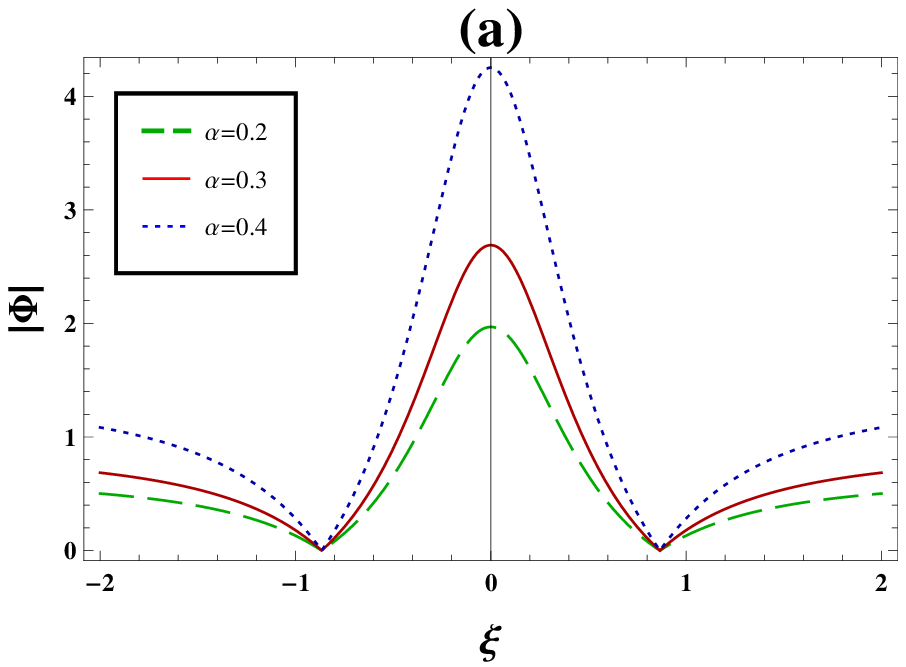}&
  \hspace{0.15in}
  \includegraphics[width=80mm]{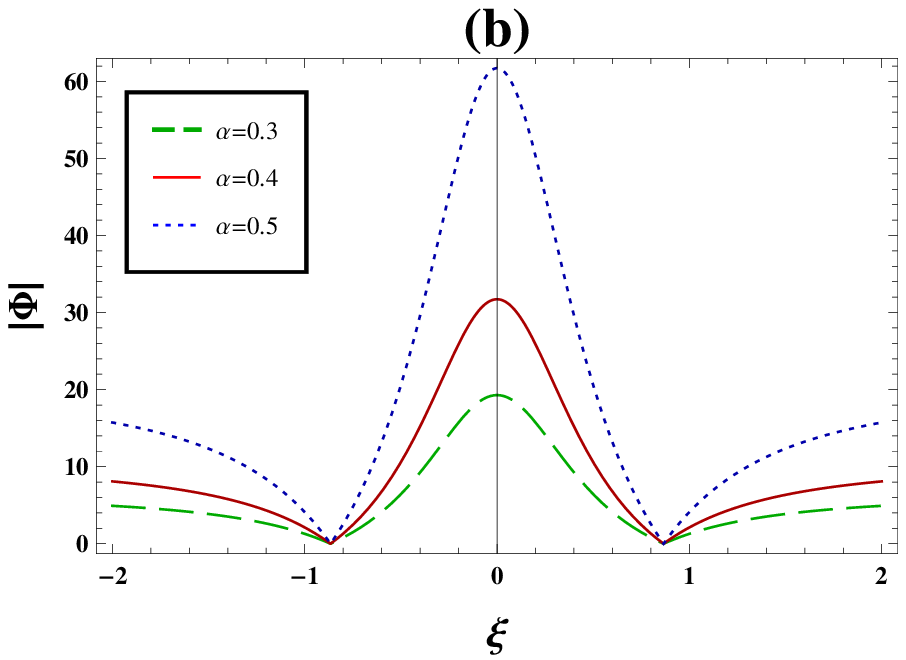}\\
  \end{tabular}
  \label{figur}\caption{Variation of $|\Phi|$ against $\xi$ for different values of $\alpha$.
  (a) For $k=0.2$ and $q=1.5$. (b) For $k=0.5,$ and $q=-0.7$.  Along with $\tau=0$.}
\end{figure*}
\begin{figure*}[htp]
  \centering
  \begin{tabular}{ccc}
  \includegraphics[width=80mm]{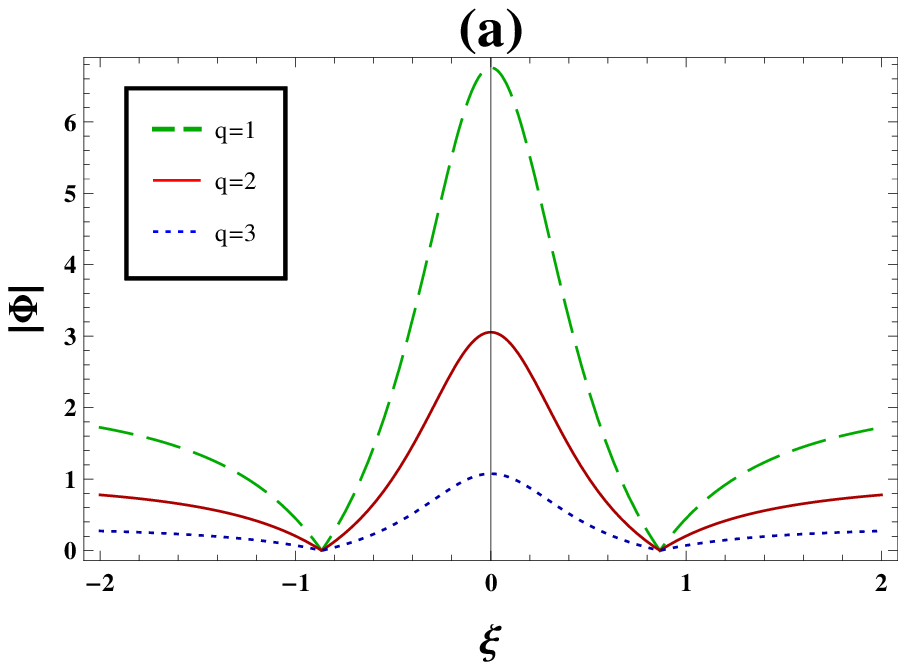}&
  \hspace{0.15in}
  \includegraphics[width=80mm]{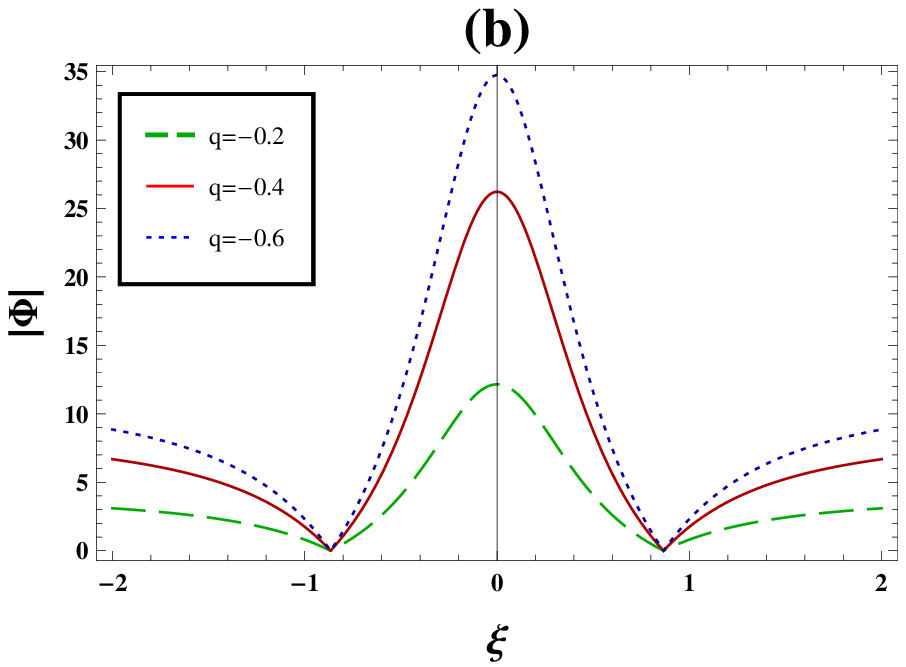}\\
  \end{tabular}
  \label{figur}\caption{Variation of $|\Phi|$ against $\xi$ for different values of $q$.
  (a) For $k=0.2$. (b) For $k=0.5$. Along with $\tau=0$.}
\end{figure*}
\section{Derivation of the NLS equation}
In order to investigate the modulation of the HIAWs in our considered plasma system,
we will derive the NLS equation by employing the reductive perturbation method. The
independent variables are stretched as
\begin{eqnarray}
&&\hspace*{-4.0cm}\xi={\epsilon}(x  - v_gt),~~~\tau={\epsilon}^2 t,\label{eq6}
\end{eqnarray}
where $v_g$ is the group velocity of the wave and $\epsilon$ is a small parameter.
Then we can write a general expression for the dependent variables as
\begin{eqnarray}
&&\hspace*{-0.2cm}n_1=1 +\sum_{m=1}^{\infty}\epsilon^{(m)}\sum_{l=-\infty}^{\infty}n_{1l}^{(m)}(\xi,\tau) ~\mbox{exp}[il(kx-\omega t)], \nonumber\\
&&\hspace*{-0.2cm}u_1=\sum_{m=1}^{\infty}\epsilon^{(m)}\sum_{l=-\infty}^{\infty}u_{1l}^{(m)}(\xi,\tau) ~\mbox{exp}[il(kx-\omega t)], \nonumber\\
&&\hspace*{-0.2cm}n_2=1 +\sum_{m=1}^{\infty}\epsilon^{(m)}\sum_{l=-\infty}^{\infty}n_{2l}^{(m)}(\xi,\tau) ~\mbox{exp}[il(kx-\omega t)], \nonumber\\
&&\hspace*{-0.2cm}u_2=\sum_{m=1}^{\infty}\epsilon^{(m)}\sum_{l=-\infty}^{\infty}u_{2l}^{(m)}(\xi,\tau) ~\mbox{exp}[il(kx-\omega t)], \nonumber\\
&&\hspace*{-0.1cm}\phi=\sum_{m=1}^{\infty}\epsilon^{(m)}\sum_{l=-\infty}^{\infty} \phi_{l}^{(m)}(\xi,\tau) ~\mbox{exp}[il(kx-\omega t)],\label{eq11}
\end{eqnarray}
where k and $\omega$ are real variables representing the carrier wave number
and frequency respectively. Since $n_1,~u_1,~n_2,~u_2$, and $\phi$
must be real, all elements in Eq. $(9)$ satisfy the reality
condition $ A_l^{(m)}= A_{-l}^{(m)^*}$, where the asterisk denotes
the complex conjugate. The derivative operators in the above equations are treated as follows:
\begin{eqnarray}
&&\hspace*{-3.7cm}\frac{\partial}{\partial t}\rightarrow\frac{\partial}{\partial t}-\epsilon v_g \frac{\partial}{\partial\xi}+\epsilon^2\frac{\partial}{\partial\tau}, \nonumber\\
&&\hspace*{-3.7cm}\frac{\partial}{\partial x}\rightarrow\frac{\partial}{\partial x}+\epsilon\frac{\partial}{\partial\xi}. \label{eq10}
\end{eqnarray}
Substituting Eqs. $(9)$ and $(10)$ into Eqs. $(1)-(4)$ and $(7)$, then the first-order approximation $(m=1)$ with the first harmonic $(l=1)$
yields the following relation
\begin{eqnarray}
&&\hspace*{-0.6cm}-i\omega n_{11}^{(1)}+iku_{11}^{(1)}=0,~~-i\omega u_{11}^{(1)}+ik\phi_1^{(1)}=0,\nonumber\\
&&\hspace*{-0.6cm}-i\omega n_{21}^{(1)}+ikn_{21}^{(1)}=0,~~-ik\alpha\phi_1^{(1)}-i\omega u_{21}^{(1)}=0,\nonumber\\
&&\hspace*{-0.6cm} n_{11}^{(1)}-k^2\phi_1^{(1)}-\gamma_1\phi_1^{(1)}-\lambda n_{21}^{(1)}=0.\label{eq14}
\end{eqnarray}
The solution for the first harmonics read as
\begin{eqnarray}
&&\hspace*{-2.0cm} n_{11}^{(1)}=\frac{k^2}{\omega^2}\phi_1^{(1)},~~~~~~~~ u_{11}^{(1)}=\frac{k}{\omega}\phi_1^{(1)} ,\nonumber\\
&&\hspace*{-2.0cm} n_{21}^{(1)}=-\frac{\alpha k^2}{\omega^2}\phi_1^{(1)},~~~~u_{21}^{(1)}=-\frac{\alpha k}{\omega}\phi_1^{(1)},\label{eq12}
\end{eqnarray}
 we thus obtain the dispersion relation for HIAWs
\begin{eqnarray}
&&\hspace*{-5.0cm} \omega^2=\frac{k^2(1+\alpha\lambda) }{k^2+\gamma_1}.\label{eq12}
\end{eqnarray}
The second-order when $(m=2)$ reduced equations with $(l=1)$ are
\begin{eqnarray}
&&\hspace*{-2.2cm}n_{11}^{(2)}=\frac{k^2}{\omega^2}\phi_1^{(2)}+\frac{2ik(v_g k-\omega)}{\omega^3} \frac{\partial \phi_1^{(1)}}{\partial\xi},\nonumber\\
&&\hspace*{-2.2cm}u_{11}^{(2)}=\frac{k}{\omega }\phi_1^{(2)}+\frac{i(v_g k-\omega)}{\omega^2} \frac{\partial \phi_1^{(1)}}{\partial\xi},\nonumber\\
&&\hspace*{-2.2cm}n_{21}^{(2)}=-\frac{\alpha k^2}{\omega^2}\phi_1^{(2)}-\frac{2ik \alpha(v_g k-\omega)}{\omega^3} \frac{\partial \phi_1^{(1)}}{\partial\xi},\nonumber\\
&&\hspace*{-2.2cm}u_{21}^{(2)}=-\frac{\alpha k}{\omega }\phi_1^{(2)}-\frac{i\alpha(v_g k-\omega)}{\omega^2} \frac{\partial \phi_1^{(1)}}{\partial\xi},\label{eq16}
\end{eqnarray}
whereas the second-order approximation $(m=2)$ with the first harmonic $(l=1)$ gives
\begin{eqnarray}
&&\hspace*{-5.0cm}v_g=\frac{\omega(1+\alpha\lambda-\omega^2) }{k(1+\alpha\lambda)}.\label{eq17}
\end{eqnarray}
The amplitude of the second-order harmonics are found to be proportional to $|\phi_1^{(1)}|^2$
\begin{eqnarray}
&&\hspace*{-2.3cm}n_{12}^{(2)}=C_1|\phi_1^{(1)}|^2,~~~~~~~~n_{10}^{(2)}=C_6 |\phi_1^{(1)}|^2,\nonumber\\
&&\hspace*{-2.3cm}u_{12}^{(2)}=C_2|\phi_1^{(1)}|^2,~~~~~~~~u_{10}^{(2)}=C_7|\phi_1^{(1)}|^2,\nonumber\\
&&\hspace*{-2.3cm}n_{22}^{(2)}=C_3|\phi_1^{(1)}|^2,~~~~~~~~n_{20}^{(2)}=C_8|\phi_1^{(1)}|^2,\nonumber\\
&&\hspace*{-2.3cm}u_{22}^{(2)}=C_4 |\phi_1^{(1)}|^2,~~~~~~~~u_{20}^{(2)}=C_9 |\phi_1^{(1)}|^2,\nonumber\\
&&\hspace*{-2.3cm}\phi_2^{(2)}=C_5|\phi_1^{(1)}|^2,~~~~~~~~\phi_0^{(2)}=C_{10}|\phi_1^{(1)}|^2,\label{eq16}
\end{eqnarray}
where
\begin{eqnarray}
&&\hspace*{-0.1cm}C_1=\frac{3k^4+2C_5~\omega^2k^2}{2\omega^4}, \nonumber\\
&&\hspace*{-0.1cm}C_2=\frac{k^3+2kC_5\omega^2}{2\omega^3} \nonumber\\
&&\hspace*{-0.1cm}C_3=\frac{3\alpha^2k^4-2\alpha k^2 \omega^2 C_5}{2\omega^4},\nonumber\\
&&\hspace*{-0.1cm}C_4=\frac{\alpha^2 k^3-2\alpha k C_5\omega^2 }{2\omega^3},  \nonumber\\
&&\hspace*{-0.1cm}C_5=\frac{2\gamma_2\omega^4+3\lambda \alpha^2k^4-3k^4}{2\omega^2k^2-2\omega^4(4k^2+\gamma_1)+2\alpha\lambda \omega^2 k^2},\nonumber\\
&&\hspace*{-0.1cm}C_6=\frac{2v_g k^3+\omega k^2+C_{10}\omega^3 }{v^2_g\omega^3}, \nonumber\\
&&\hspace*{-0.1cm}C_7=\frac{k^2+C_{10}\omega^2}{v_g\omega^2}, \nonumber\\
&&\hspace*{-0.1cm}C_8=\frac{2v_g \alpha^2 k^3+\omega \alpha^2 k^2-\alpha C_{10}\omega^3 }{v^2_g \omega^3}, \nonumber\\
&&\hspace*{-0.1cm}C_9=\frac{\alpha^2 k^2-\alpha C_{10} \omega^2 }{v_g \omega^2},  \nonumber\\
&&\hspace*{-0.1cm}C_{10}=\frac{2\gamma_2 v^2_g \omega^3+2\lambda v_g \alpha^2k^3+\omega \lambda \alpha^2 k^2-2v_g k^3-\omega k^2}{\omega^3(1+\alpha\lambda-\gamma_1 v^2_g)}.\nonumber\
\end{eqnarray}
Finally, the third harmonic modes $(m=3)$ and $(l=1)$ and  with the help of  Eqs. $(12) - (16)$,
give a system of equations, which can be reduced to the following NLS equation:
\begin{eqnarray}
&&\hspace*{-4.1cm}i\frac{\partial \Phi}{\partial \tau}+P\frac{\partial^2 \Phi}{\partial \xi^2}+Q|\Phi|^2\Phi=0, \label{eq24}
\end{eqnarray}
where $\Phi=\phi_1^{(1)}$ for simplicity. The dispersion coefficient $P$ is
\begin{eqnarray}
&&\hspace*{-5.2cm}P=\frac{3}{2}\left(\frac{v_g}{\omega}-\frac{1}{k}\right)v_g,\nonumber\
\end{eqnarray}
and the nonlinear coefficient $Q$ is given by
\begin{eqnarray}
&&\hspace*{-0.3cm} Q=\frac{\omega^3}{2k^2(1+\alpha\lambda)}\left[2\gamma_2(C_5+C_{10})+3\gamma_3-\frac{k^2(C_1+C_6)}{\omega^2} \right.\nonumber\\
&&\hspace*{-0.3cm}\left.-\frac{2k^3(C_2+C_7)}{\omega^3} -\frac{\alpha\lambda k^2(C_3+C_8)}{\omega^2}-\frac{2\alpha \lambda k^3(C_4+C_9)}{\omega^3}\right].\nonumber\
\end{eqnarray}
\section{stability and rogue waves}
The nonlinear evolution of the HIAWs typically depends on the coefficients of dispersion
term $P$ and nonlinear term $Q$ which are function of the various plasma parameters such
as $\alpha,~\sigma,~\mu_e,~\mu_p$, and $q$. Thus, these  plasma parameters are significantly
controlled the stability conditions of the HIAWs. If $PQ<0$, HIAWs are modulationally stable
and for the case $PQ>0$, HIAWs are modulationally unstable against external perturbations
\cite{Kourakis2005,Sukla2002,Schamel2002,Fedele2002}  and simultaneously
when $PQ>0$ and ${k_{MI}}<k_c$, the growth rate ($\Gamma_g$) of  MI is
given \cite{Shalini2015}  by
\begin{eqnarray}
&&\hspace*{-4.5cm}\Gamma_g=|P|~{k^2_{MI}}\sqrt{\frac{k^2_{c}}{k^2_{MI}}-1}.\label{eq112}
\end{eqnarray}
Here ${k_{MI}}$ is the perturbation wave number and the critical value of the wave number of modulation
$k_c=\sqrt{2Q{|\Phi_o|}^2/P}$, where $\Phi_o$ is the amplitude of the carrier waves.
Hence the maximum value $\Gamma_{g(max)}$ of $\Gamma_g$ is obtained at ${k_{MI}}=k_c/\sqrt{2}$
and is given by $\Gamma_{g(max)}=|Q||\Phi_0|^2$.

Therefore, we have investigated the stability of the  profile by depicting the ratio of $P/Q$ versus
$k$ for different plasma parameters. When  $P/Q<0$, HIAWs are modulationally stable, while  $P/Q>0$, HIAWs will be modulationally  unstable
against external perturbations.  When $P/Q\rightarrow\pm\infty$, the corresponding value of $k(=k_c)$
is called critical or threshold wave number for the onset of MI. Figure $1$ shows the
variations of $P/Q$ with $k$ for possible three ranges $-1<q<0$, $0<q<1$, and $q>1$ of
the nonextensive parameter $q$. It is observed from both Figs. $1(a)$  and $1(b)$
that for small $k$  there is stable region (HIAWs are modulationally stable
and dark envelope solitons exist), where unstable region (HIAWs are modulationally
unstable and bright envelope solitons exist) is found for large $k$. The critical value
$k_c$ is increases (decreases) with an increase in the value of nonextensive
parameter $q$ for the ranges $q>0$ ($q<0$) which can be observed from  Figs. $1(a)$ and  $1(b)$,
respectively. So the range of nonextensive parameter  (greater than or less than zero) plays a
vital role for controlling the stability of the HIAWs.

We are now investigating the effects of the ratio of electron to light positive ion number density
(via $\mu_e$) on stability conditions of HIAWs which are depicted in Figs. $2(a)$ and $2(b)$.
Nonextensive electrons  play  a fascinating role to control the stability of HIAWs structure.
Within the range $q>0$, increasing values of electron number density, the critical value
$k_c$ is shifted to the higher values that means the instability domain is occurred
at the higher values of $k$ in Fig. $2(a)$. Similar behaviour is observed  for the case  $q<0$,
the critical value $k_c$ is occurred at higher wave number with increasing of electrons
number density in Fig. $2(b)$. So excess number of electrons are provided
more restoring force which  maximize the stability region without depending on the  values of nonextensive parameter.

The effects of  temperature parameter  (via $\sigma)$ on stable and unstable regions
can be shown from   Figs. $3(a)$ and $3(b)$, respectively. The variation of  $P/Q$ with $k$  for
different values of $\sigma$ is depicted in  Figs. $3(a)$ and $3(b)$ . It is observed
that the critical value $(k_c)$  is shifted towards higher (lower) values as electron
temperature increases for the range of $q>0$ ($q<0$). The range of
nonextensive parameter change the order of the variation of critical value $(k_c)$.
Thus, electron temperature plays a crucial role to change the stability of the wave packets.

To highlight the effects of the ratio of positron to light positive ion number density (via $\mu_p$) on the growth rate of MI is
depicted in Fig. $4$. The growth rate of the MI is so much sensitive to change the
values of $\mu_p$. It is observed from Figs. $4(a)$ and $4(b)$ that an increase
of $\mu_p$ value the maximum growth rate is increased (decreased) for $q>0~(q<0)$,
respectively. So the nonextensive positrons (via $\mu_p$) can be recognized to enhance or
suppress of the MI growth rate of our considered plasma model.

The  rogue wave (rational solution) of the NLS Eq. $(17)$ in the unstable
region ($PQ>0$) can be written \cite{Ankiewiez2009,Abdelwahed2016} as
\begin{eqnarray}
&&\hspace*{-0.9cm}\Phi(\xi,\tau)=\sqrt{\frac{2P}{Q}} \left[\frac{4(1+4iP\tau)}{1+16P^2\tau^2+4\xi^2}-1 \right]\mbox{exp}(i2P\tau).\label{eq12}
\end{eqnarray}
The solution $(19)$ anticipates the concentration of the HIAWs energy into a small
region that is caused by the nonlinear behavior of the plasma medium (see Figs. $5$ and $6$).
The HIARWs are so much sensitive to change any plasma parameter.
The effect of the ratio of light positive ion mass to heavy negative ion mass (via $\alpha$) on the HIARWs
is shown in Fig. $5(a)$ and $5(b)$. For the range $q>0$~$(q<0)$, the amplitude and width of the HIARWs
are increased with the increase of $\alpha$ which can be seen in Figs. $5(a)$ and $5(b)$. However, the HIARWs amplitude
decreases with the increase of the heavy negative ion mass (via $\alpha$). Physically,
the increasing of the heavy negative ions masses lead to dissipate the energy from the system and reduce
the nonlinearity that makes the HIARWs amplitude shorter. Finally, changing direction of HIARWs
amplitude and width  are remain invariant, irrespective of $q$ values. Exact similar fashion
can be observed in Fig. $6$, which is depicted against $q$. For the range $q>0$~$(q<0)$, the amplitude
and width of HIARWs is decreased (decreased) with the increase of $q$ values.

It can be deduced from Figs. $2$ and $5$ that  the direction of the variation of  $k_c$ and
wave profile (hight and width) is totally independent on the nonextensive parameter
ranges.  For the ranges of $q>0$ and $q<0$, with an increase of the ratio of electron
to positive light ion number density (via $\mu_e$) and the ratio of light positive ion
to heavy negative ion mass (via $\alpha$), the values of $k_c$ and wave profile (hight and width) are
increased which can be observed from Figs. $2$ and $5$, respectively.
\section{Conclusion}
In summary a NLS  equation has been derived to describe the small-amplitude HIAWs
in an unmagnetized multi-component plasma consisting of inertial light positive
ions, heavy negative ions, as well as inertialess nonextensive electrons and
positrons. The analysis of the NLS equation reveals the existence of both stable
and unstable region. The critical value  $k_c$ which determined the stability/instability
region of HIAWs, is totally depend on various plasma parameters such as nonextensive
parameter, electron number density, and electron temperature. Every plasma
parameters plays a vital role to change the critical value $k_c$.
The MI of HIAWs can also lead to the formation of rogue waves in the unstable region in which a
large amount of energy is concentrated in relatively small area in space and time.
The hight and width of rogue waves are greatly depended on the ratio of light positive
ion to heavy negative ion mass which is observed in our present analysis.
A  number of observations clearly disclosed the existence of nonextensive electrons
and positrons in various natural space environment and laboratory plasmas. We
are optimistic that the finding of our present  investigation should be useful
to understand the nonlinear phenomena in both  space (cometary comae and
earth's ionosphere) and laboratory plasmas (laser plasma)
which containing of inertial light positive ions and  heavy  negative ions, as
well as inertialess nonextensive electrons and positrons.
\section*{Acknowledgements}
N. A. Chowdhury is grateful to the Bangladesh Ministry of Science and Technology for
awarding the National Science and Technology (NST) Fellowship.


\begin{thebibliography}{99}

\bibitem{Misner1973} W. Misner, K. S. Thorne, and J. I. Wheeler, \textit{Gravitation} (Freeman, San Francisco, 1973).

\bibitem{Miller1987} H. R. Miller and P. J. Witter, {\em Active Galactic Nuclei} (Springer, Berlin, 1987), p. 202.

\bibitem{Michel1982} F. C. Michel, Rev. Mod. Phys. {\bf54}, 1 (1982).

\bibitem{Michel1991} F. C. Michel, \textit{Theory of Neutron Star Magnetosphere} (Chicago University Press, Chicago, 1991).

\bibitem{Shukla1984} P. K. Shukla, M. Y. Yu, and N. L. Tsintsade, Phys. Fluids {\bf27}, 327 (1984).

\bibitem{Shukla1986} P. K. Shukla, N. N. Rao, M. Y. Yu, and N. L. Tsintsade, Phys. Rep. {\bf138}, 1 (1986).

\bibitem{Surko1990} C. M. Surko and T. J. Murphy, Phys. Fluids {\bf2}, 1372 (1990).

\bibitem{Chaizy1991} P. H. Chaizy, H. Reme, J. A. Sauvaud, C. Duston, R. P. Lin, D. E. Larson,
D. L. Mitchell, K. A. Anderson, C. W. Carlson, A. Korth, and D. A. Mendis, Nature (London) {\bf349}, 393 (1991).

\bibitem{Massey1976} H. Massey, \textit{Negative Ions}, 3rd edn. (Cambridge univesity press, Cambridge, 1976).

\bibitem{Vladimirov2003} S. V. Vladimirov, K. Ostrikov, M. Y. Yu, and G. E. Morfill, Phys. Rev. E {\bf67}, 036406 (2003).

\bibitem{Mamun2003} A. A. Mamun and P. K. Shukla, Phys. Plasmas {\bf10}, 1518 (2003).

\bibitem{Djebli2003} M. Djebli, Phys. Plasmas {\bf10}, 4910 (2003).

\bibitem{Bacal1979} M. Bacal and G. W. Hamilton, Phys. Rev. Lett. {\bf42}, 1538 (1979).

\bibitem{Gottscho1986} R. A. Gottscho and C. E. Gaebe,  IEEE Trans. Plasma Sci. {\bf14}, 92 (1986).

\bibitem{Ichiki2002} R. Ichiki, S. Yoshimura, T. Watanabe, Y. Nakamura, and Y. Kawai, Phys. Plasmas {\bf9}, 4481 (2002).

\bibitem{Kourakis2006} I. Kourakis, A. Esfandyari-Khalejahi, M. Mehdipoor, and P. K. Shukla, Phys. Plasmas {\bf13}, 052117 (2006).

\bibitem{Esfandyari-Khalejahi2006} A. Esfandyari-Khalejahi, I. Kourakis, M. Mehdipoor, and P. K. Shukla, J. Phys. {\bf39}, 13817 (2006).

\bibitem{Maksivonic1997} M. Maksivonic, V. Pierrard, and P. Riley, Geophys. Res. Lett. {\bf24}, 1511 (1997).

\bibitem{Leubner2004} M. P. Leubner et al., Phys. Plasmas {\bf11}, 1308 (2004).

\bibitem{Christon2012} S. P. Christon, Phys. Plasmas {\bf93}, 2562 (2012).

\bibitem{Renyi1955} A. Renyi, Acta Math. Acad. Sci. Hung. {\bf6}, 285 (1955).

\bibitem{Tsallis1988} C. Tsallis, J. Stat. Phys. {\bf52}, 479 (1988).

\bibitem{Silva1998} R. Silva Jr., A. R. Plastino, and J. A. S. Lima, Phys. Lett. A {\bf249}, 401 (1998).

\bibitem{Plastino1993} A. R. Plastino and A. Plastino, Phys. Lett. A {\bf174}, 384 (1993).

\bibitem{Gervino2012} G. Gervino, A. Lavagno, and D. Pigato, Cent. Eur. J. Phys. {\bf10}, 594 (2012).

\bibitem{Feron2008} C. Feron and J. Hjorth, Phys. Rev. E {\bf77}, 022106 (2008).

\bibitem{Asbridge1968} J. R. Asbridge, S. J. Bame, and I. B. Strong, J. Geophys. Res. {\bf73}, 5777, (1968).

\bibitem{Krimigis1983} S. M. Krimigis, J. F. Carbary, E. P. Keath, T. P. Armstrong, L. J. Lanzerotti, and G. Gloeckler, J. Geophys. Res. {\bf88}, 8871 (1983).

\bibitem{Vladimirov2004} S. V. Vladimirov and K. J. Ostrikov, Phys. Rep. {\bf393}, 175 (2004).

\bibitem{Kharif2009} C. Kharif, E. Pelinovsky, and A. Slunyaev, \textit{Rogue waves in the Ocean} (Springer-Verlag, Berlin, 2009).

\bibitem{Yan2010} Z. Yan, Commun. Theor. Phys. {\bf54}, 947 (2010).

\bibitem{Ganshin2008} A. N. Ganshin, V. B. Efimov, G. V. Kolmakov, L. P. Mezhov-Deglin, and P. V. E. McClintock, Phys. Rev. Lett. {\bf101}, 065303 (2008).

\bibitem{Eslami2011} P. Eslami, M. Mottaghizadeh, and H. R. Pakzad, Phys. Plasmas {\bf18}, 102313 (2011).

\bibitem{Bacha2012} M. Bacha, S. Boukhalfa, and M. Tribeche, Astrophys. Space Sci. {\bf341}, 591 (2012).

\bibitem{El-Labany2011} S. K. El-Labany, W. M. Moslem, N. A. El-Bedwehy, R. Sabry, and H. N. Abd El-Razek, Astrophys. Space Sci. {\bf338}, 3 (2011).

\bibitem{Jannat2016} N. Jannat, M. Ferdousi, and A. A. Mamun, Plasma Phys. Rep. {\bf42}, 678 (2016).

\bibitem{Kourakis2005} I. Kourakis and P.K. Shukla, Nonlinear Proc. Geophys.  {\bf12}, 407 (2005).

\bibitem{Sukla2002} R. Fedele, H. Schamel, and P.K. Shukla, Phys. Scr.  {\bf98}, 18 (2002).

\bibitem{Schamel2002} R. Fedele and H. Schamel, Eur. Phys. J. B  {\bf27}, 313 (2002).

\bibitem{Fedele2002} R. Fedele, Phys. Scr. {\bf65}, 502 (2002).

\bibitem{Shalini2015} Shalini, N. S. Saini, and A. P. Misra, Phys. Plasmas {\bf22}, 092124 (2015).

\bibitem{Ankiewiez2009} A. Anikiewicz, N. Devine, and N. Akhmediev, Phys. Lett. A {\bf373}, 3997 (2009).

\bibitem{Abdelwahed2016} H. G. Abdelwahed, E. k. El-Shewy, M. A. Zahran, and S. A. Elwakil, Phys. Plasmas {\bf23}, 022102 (2016).

\end{thebibliography}
\end{document}